\documentclass[aps, prl, twocolumn,amsmath,amssymb,floatfix,showpacs]{revtex4}
\usepackage{amsmath,amssymb,amsfonts,mathrsfs}
\usepackage{color}
\usepackage{epsfig}
\usepackage{psfrag}
\usepackage{graphicx}
\usepackage{color}

\begin{document}

\title{Majorana Fermions in superconducting 1D systems having periodic,
quasiperiodic, and disordered potentials}

\author{Wade DeGottardi$^1$, Diptiman Sen$^{2}$, and Smitha Vishveshwara$^1$}
\affiliation{\small{
$^1$Department of Physics, University of Illinois at Urbana-Champaign,
1110 W.\ Green St.\ , Urbana, IL 61801-3080, USA\\
$^2$Centre for High Energy Physics, Indian Institute of Science, Bangalore
560012, India}}
\pacs{03.65.Vf, 71.10.Pm}

\date{\today}

\begin{abstract}{We present a unified study of the effect of periodic,
quasiperiodic and disordered potentials on topological phases that are
characterized by Majorana end modes in 1D $p$-wave superconducting systems.
We define a topological invariant derived from the equations of motion for
Majorana modes and, as our first application, employ it to characterize the
phase diagram for simple periodic structures. Our general result is a
relation between the topological invariant and the normal state localization
length. This link allows us to leverage the considerable literature on
localization physics and obtain the topological phase diagrams and their
salient features for quasiperiodic and disordered systems for the entire
region of parameter space.}
\end{abstract}

\maketitle

\begin{center}
\begin{figure}
\includegraphics[bb=0 0 640 480,width=9.5cm]{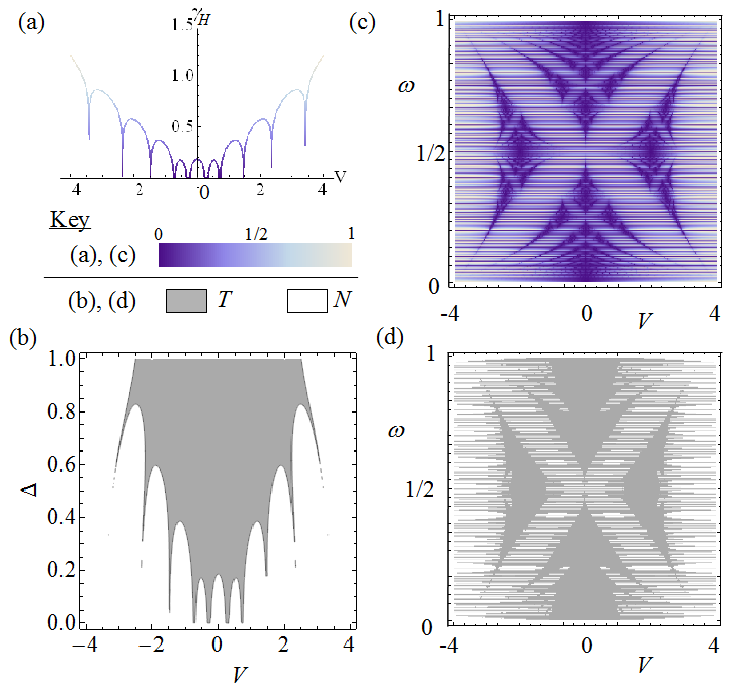}
\caption{(a) The Lyapunov exponent $\gamma_H$ of the normal state ($\Delta =0$)
for $\mu_n = V + 2 \cos \left( 2 \pi \omega n \right)$ with $\omega = 1/10$.
(b) Topological phase boundary showing the merging of the topological regions
(T) as described in the text. For $\Delta \ll 1$, there are are 10 distinct
regions of non-trivial topology (T). At $\Delta = 0.2$, the four central
regions have merged to form a single region. (c) A colorscale plot of
$\gamma_H$ as a function of $V$ and $\omega$. Darker regions correspond to
lower values of $\gamma_H$. Striations show the spectrum's sensitivity to $q$
for $\omega=p/q$. (d) The topological phase diagram ($\Delta = 0.2$)
reflects the geometry of the low-lying values of $\gamma_H$ in (c).}
\label{fig:butterfly}
\end{figure}
\end{center}

\emph{Introduction -- } Recent claims of the detection of Majorana fermions
in semiconducting/superconducting heterostructures have stirred new
excitement leading to several avenues of inquiry~\cite{kouwenhoven,alicea}.
A major concern in these effectively spinless $p$-wave superconducting wires
is the role of spatially varying potentials, be they externally
applied or due to disorder. The latter has in fact been a rich, active topic
of study for over a decade in terms of delocalization-localization physics in
one-dimensional (1D) class D systems~\cite{classD,motrunich}.
The pioneering work of Ref.~\cite{motrunich} specifically explored the
conditions for the existence
of Majorana boundary modes in these disordered systems, arguing that a
finite amount of superconductivity is required to drive the system into
such a topological phase. More recently, several works
have further investigated this aspect, primarily exploring either weak
(or slowly varying) disorder or a special point in parameter space
corresponding to the disordered quantum Ising chain~\cite{fisher,shivamoggi}.
In this Letter, we perform a cohesive study of the topological
phase diagram for a range of potential landscapes on a lattice, extending
and unifying work on periodic~\cite{ladder,period},
incommensurate~\cite{incommensurate}, and disordered
potentials~\cite{motrunich,brouwer,dassarma}.
\begin{center}
\begin{figure}
\includegraphics[bb=0 0 380 380,width=9cm]{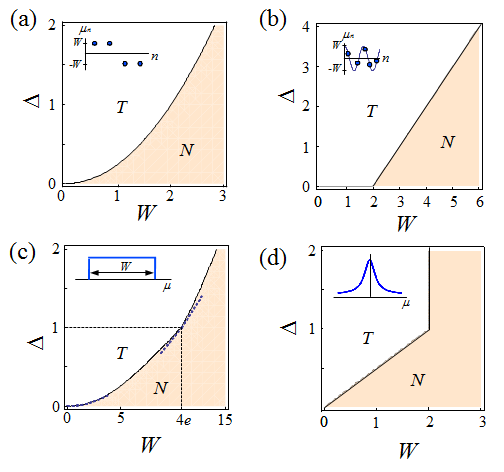}
\caption{Topological phase diagrams (T - topological; N - non-topological)
as a function of superconducting
gap, $\Delta$, and potential strength, $W$, for (a) a periodic potential
having the pattern $(W,W,-W,-W)$, (b) a quasiperiodic potential
$W\cos(2\pi\omega n)$ for any irrational $\omega$, (c) box-distributed
(uniform) disorder, and (d) Lorentzian distributed disorder.}
\label{fig:disorder}
\end{figure}
\end{center}
Our central observation links the normal state localization properties to the
strength of the superconducting pairing required to engender an end Majorana
mode, thus enabling us to obtain the topological phase diagram for a variety
of situations (see Fig.~\ref{fig:disorder}). We define a topological phase
(T-phase) to be one in which there are zero-energy Majorana modes at the ends
of an infinitely long system. Specifically, we define a topological invariant
based on the decay length of the end mode~\cite{ladder} and provide a
relationship between this length and the corresponding zero-energy
localization length of the normal system. Unlike in the uniform case, we
find that even in the simplest case of periodic potentials, the T-phase
requires a finite amount of superconductivity,
which thus acts as a new knob to access the phase. In the quasiperiodic case,
the topological boundaries reflect the morphology of the fractal patterns
exhibited by the normal state energy spectrum (see Fig.~\ref{fig:butterfly}).

For disordered systems, our observation proves powerful in that it allows
us to leverage the vast body of
literature on normal state localization physics to identify topological
properties of the disordered superconductor. The topological regions arise as
a result of the competition between the localizing effects of disorder and the
superconducting pairing which tends to spatially separate the Majorana
fermions composing a Dirac state. Our analyses provide a mapping for the
phase boundary between strong and weak disorder limits and reproduce the
exact form for the intermediate point corresponding to the random field
transverse Ising model. We present two representative cases of our analyses
for the disordered phase diagram in Fig.~\ref{fig:disorder}(c-d).

\emph{Model and Topological features --} Turning to our starting point, we
model the superconducting wire as a 1D tight-binding system of spinless
electrons exhibiting $p$-wave superconductivity described by the
$\mathcal{N}$-site Hamiltonian~\cite{kitaev}
\begin{eqnarray} H = \sum_{n = 1}^{\mathcal{N}} \Big[ &-& t \left(f_n^\dagger
f_{n+1}^{\phantom\dagger} + f_{n+1}^\dagger f_n^{\phantom\dagger} \right) +
\Delta \Big( f_n^{\phantom\dagger} f_{n+1}^{\phantom\dagger} \nonumber \\
&+& f_{n+1}^\dagger f_n^\dagger \ \Big) - \mu_n \left(f_n^\dagger
f_n^{\phantom\dagger} - 1/2 \right) \Big] , \label{eq:hamdisorder}
\end{eqnarray}
where $t$ is the nearest-neighbor hopping amplitude, $\Delta$ is the
superconducting gap function (taken to be real), and we will eventually take
$\mathcal{N} \to \infty$. The various cases of periodic, quasiperiodic, and
disordered potentials are encoded in the local on-site chemical potential
$\mu_n$ and are characterized by a typical potential strength $W$. The Dirac
fermion $f_n$ can be expressed in terms of Majorana fermions, $f_n =
(a_n + i b_n)/2$, which are Hermitian operators satisfying the
anticommutation rules $\{a_n, a_m \} = \{b_n, b_m \} =
2 \delta_{n,m}$ and $\{a_n, b_m \} = 0$.

We now construct a topological invariant that links topology to the
eigenvalue structure of zero-energy boundary modes described by the
Hamiltonian in Eq.~(\ref{eq:hamdisorder}). Specifically, the end Majorana
modes that decay into the bulk can be represented by the operators $Q_a =
\sum_n \alpha_n a_n, ~Q_b = \sum_n \beta_n b_n$, where the wave function
$\alpha_n$ obey the zero-energy equations of motion derived from
Eq.~(\ref{eq:hamdisorder}). As detailed in Ref.~\cite{ladder}, these
equations can be represented in the transfer matrix form
\begin{equation}
\label{eq:matrixA}
\left( \begin{array}{c}
\alpha_{n+1} \\
\alpha_n \end{array} \right) =~A_n \left( \begin{array}{c}
\alpha_n \\
\alpha_{n-1} \end{array} \right),~\mbox{where}~
A_n = \left( \begin{array}{cc}
\frac{\mu_n }{\Delta + t} & \frac{\Delta - t}{\Delta + t} \\
1 & 0 \end{array} \right). \end{equation}
Since the $A_n$ may be taken as functions of $\mu_n / t$ and $\Delta / t$,
we set $t = 1$. A similar expression holds for the $\beta_n$.

The existence of end Majorana modes requires the $\alpha_n$ (or $\beta_n$) to
be normalizable. We denote the number of eigenvalues of the transfer matrix
$\mathcal{A}(W,\Delta) \equiv \prod_{n = 1}^{\mathcal{N}} A_n$
with magnitude less than 1 by $n_f$. For $n_f = 0,2$, the system is
topological, whereas for $n_f = 1$
it is non-topological~\cite{ladder}. Since the topology of the system depends
only on the magnitude of $\Delta$, we take $\Delta$ to be positive;
hence $|\det~\mathcal{A}| < 1$. Then the two eigenvalues of $\mathcal{A}$
obey $|\lambda_1 \lambda_2| < 1$. Therefore, for $|\lambda_1|<|\lambda_2|$,
we have $|\lambda_1| < 1$ and $n_f$ is completely determined by the larger
eigenvalue $\lambda_2$. We thus define a topological invariant
\begin{equation}
\nu ~=~ -~ (-1)^{n_f} ~=~ \mbox{sgn} \left( \ln |\lambda_2| \right),
\label{eq:TIinvariant} \end{equation}
which can also be expressed as $\nu = - \mbox{sgn} \left( f(1) f(-1) \right)$,
where $f(z) = \det \left(I - \mathcal{A} z \right)$ is the characteristic
polynomial of $\mathcal{A}$~\cite{ladder}. From the conditions on $n_f$, we
see that $\nu=-1$ for the T-phase and $\nu=1$ for the non-topological phase (N-phase).

\emph{Uniform and Periodic Potentials -- }
As a simple application of our topological invariant, we consider periodic
patterns in the sign of the chemical potential (detailed in
Ref.~\cite{ladder}).
Table I presents the conditions for T-phases for some select patterns. The
comparison of the phase diagrams for the uniform case and simple periodic
potentials highlights unusual aspects of the former's phase diagram. In
particular, the system is topological for $|W/t| < 2$ for any $\Delta \neq 0$.
This stems from the fact that the system has a bulk gap at $E=0$ as long as
$W$ lies in the range $[-2t,2t]$~\cite{kitaev}. In contrast, non-uniform
potentials tend to open a bulk gap at $E = 0$ with a size that grows with
increasing $W$. This leads to a phase boundary which generically has
$\frac{d \Delta}{dW} > 0$~\cite{beenakker}. We now quantify this observation and extend it to other potential landscapes.

\begin{table}
\caption{Criteria for topological phase for a selection of periodic
potentials.}
\label{tab:sectors}
\begin{center}
\begin{tabular}{|c|c|c|}
\hline
period & pattern of $\mu_n$ & topological for \\
\hline
\hline
1 & $ \ldots,W,W, W, \ldots $ & $| W | < 2$ \\
\hline
2 & $ \ldots, W, - W,\ldots$ & $ \Delta > | W | / 2$ \\
\hline
4 & $ \ldots, W, W, W, -W,\ldots$ & $\Delta^2 > W^2/2 - 1 $ \\
4 & $ \ldots, W, W, -W, -W, \ldots $ & $ \Delta > W^2/4 $ \\
\hline
\end{tabular}
\end{center}
\end{table}

\emph{Features of the Topological Phase Diagram --} As observed in
Ref.~\cite{motrunich}, the product of transfer matrices which appears in
$\mathcal{A}$ is
strongly reminiscent of that used to determine localization properties of
the normal state Anderson disorder problem. We build on this observation by
determining the phase boundary from the normal state properties of
the system. In general, this leads to a critical amount of superconductivity
required to drive the system into a T-phase.

To this end, for $0 < \Delta < 1$, we perform a similarity transformation
$A_n = \sqrt{\delta} S \tilde{A}_n S^{-1}$ with $S = \mbox{diag} (\delta^{1/4},
1/\delta^{1/4} )$ and $\delta = \frac{1-\Delta}{1+\Delta}$. The matrices
$\tilde{A}_n$ are of the form shown in Eq.~(\ref{eq:matrixA}) with $\Delta
\to 0$ and $\mu_n \to \mu_n/\sqrt{1-\Delta^2}$. This immediately gives
\begin{equation}
\mathcal{A}(W,\Delta) =\left( \sqrt{ \frac{1-\Delta}{1+\Delta}}
\right)^\mathcal{N} S \mathcal{A}\left(W/\sqrt{1-\Delta^2},0\right) S^{-1}.
\label{eq:trans} \end{equation}
Taking the logarithm of the eigenvalues of Eq.~(\ref{eq:trans}),
the condition that $|\lambda_2| = 1$ is given by
\begin{equation} \gamma \left( \frac{W}{\sqrt{1-\Delta^2}} \right)
= \frac{1}{2} \ln \left( \frac{1+\Delta}{1-\Delta} \right),
\label{eq:phaseboundary} \end{equation}
where we have defined the Lyapunov exponent of the normal state system,
$\gamma(W) \equiv \lim _{\begin{subarray}{l} \mathcal{N} \to \infty
\end{subarray}}
\frac{1}{\mathcal{N}} \ln | \lambda_2 (W,0) |$. Eq.~(\ref{eq:phaseboundary})
thus describes the phase boundary separating the topologically trivial and
non-trivial regions of the phase diagram. This relation quantifies the
observation in~\cite{motrunich,brouwer} that in general a
critical amount of superconductivity must be applied before the system is
driven into a T-phase. For the case in which the system is metallic (i.e.,
$\gamma(W) = 0$), any non-zero $\Delta$ will give rise to a T-phase.

The form of the phase diagram for $\Delta > 1$ may be obtained by noting that
the transformation
\begin{eqnarray}
\mu_n \to \mu_n/\Delta, \Delta \to 1/\Delta~,~\mbox{and}~ P \to \tilde{P} \\
\mbox{where}~P \to \tilde{P}:~ \{\mu_n\} \to \{(-1)^n \mu_n \}, \nonumber
\label{eq:duality} \end{eqnarray}
leaves the eigenvalues of $\mathcal{A}$ unchanged for $\mathcal{N}$ even. Thus, if a
point $(W_0,\Delta_0<1)$ lies on the phase boundary of $P$, then
$(W_0/\Delta_0,1/\Delta_0)$ lies on the phase boundary of $\tilde{P}$. This
duality strongly constrains the form of the phase boundary in the cases where
the distribution is
invariant under the transformation in Eq.~(\ref{eq:duality}).

Finally, at the point $\Delta = 1$, the system maps to the well-studied
quantum Ising chain subject to a random transverse
field~\cite{fisher,shivamoggi}. The matrix $\mathcal{A}(W,1)$ has the
eigenvalues $\frac{1}{2^\mathcal{N}} \prod_{n = 1}^\mathcal{N} \mu_n $ and 0.
Eq.~(\ref{eq:TIinvariant}) reveals that the phase boundary passes through the
point for which
\begin{equation} \langle \ln | \mu_n | \rangle = \ln 2,
\label{eq:specialpoint1} \end{equation}
where $\langle \ln | \mu_n | \rangle \equiv \lim_{\mathcal{N}\to \infty}
\frac{1}{\mathcal{N}} \sum_{n=1}^\mathcal{N} \ln | \mu_n |$.
These relations allow us to obtain the superconducting topological phase
diagram for quasiperiodic and disordered potentials purely based on the normal
state localization properties.

\emph{Quasiperiodic Potentials -- } Here we consider two cases of potentials
to study periodicity that is incommensurate with the underlying lattice. In
the first instance, where $\mu_n = W \cos \left( 2 \pi \omega n \right)$ and
$\omega$ is irrational, the normal state features a well-studied
metal-insulator transition at the critical value of $W = 2$~\cite{svetlana}.
The normal state Lyapunov exponent takes the form
$\gamma_{QP}(W) = \ln \left( W / 2 \right)$ for $W > 2$ and $0$ for $0 \leq
W \leq 2$ for $\omega$ irrational~\cite{delyon,andre}.
Eq.~(\ref{eq:phaseboundary}) then predicts a T-phase for
\begin{equation} \Delta > \frac{1}{2} W - 1.
\label{eq:quasiperiodic} \end{equation}
This result holds for all values of $\Delta$ given that the transformation
$\omega \to \omega + 1/2$ yields Eq.~(\ref{eq:duality}) and that the duality
transformation, $\Delta\to 1/\Delta$ and $W \to W/\Delta$, leaves
Eq.~(\ref{eq:quasiperiodic}) invariant. Finally, Eq.~(\ref{eq:specialpoint1})
also shows that the point $(W,\Delta) = (4,1)$ lies on the phase boundary.

A second example of particular interest is the Harper potential $\mu_n = V +
2 \cos \left( 2 \pi \omega n \right)$, corresponding to the problem of an
electron hopping on a 2D square lattice with each plaquette enclosing a
magnetic flux \cite{hofstadter}. The associated normal state energy spectrum,
the celebrated Hofstadter's butterfly, has a
rich fractal structure stemming from the fact that $\omega$ can take
irrational values and that values of $\omega = p/q$ ($p$ and $q$ relatively
prime) give rise to $q$ bands separated by $q-1$ non-zero gaps
(Fig.~\ref{fig:butterfly}(a)). From Eq.~(\ref{eq:TIinvariant}), we directly
obtain the topological phase diagram for non-zero $\Delta$. As shown in
Fig.~\ref{fig:butterfly}(b), for $\Delta \ll 1$, there are $q$ topological
regions inherited from the normal state which fuse as $\Delta$ is increased.
In the $\omega-V$ parameter space, as expected from our general analysis, the
normal state properties (Fig.~\ref{fig:butterfly}(c)) directly inform the
topological phase diagram (Fig.~\ref{fig:butterfly}(d)).

\emph{Disordered Potentials -- }
We begin with our most general results for the disordered topological phase
diagram which pertain to the limits of weak and strong disorder. Consider
weak, uncorrelated disorder satisfying $\langle \mu_n \mu_{n'} \rangle = U
\delta_{n,n'}$, $\langle \mu_n \rangle = 0$.
Using the known Lyapunov exponent obtained from perturbation theory for the
normal state system~\cite{derrida,disorderreview} and
Eq.~(\ref{eq:phaseboundary}), we obtain the condition for the T-phase
\begin{equation}
\Delta > \left( \frac{\Gamma(3/4)}{\Gamma(1/4)} \right)^2 U \approx 0.114 ~U.
\label{eq:low} \end{equation}
This result may be compared to that of a continuum model based on the Dirac equation, which
gives a topologically non-trivial phase for $\Delta > \frac{1}{8} U = 0.125 U$
(see~\cite{brouwer}). For disorder distributions that are symmetric around 0,
the self-duality condition $P=\tilde{P}$ in Eq.~(\ref{eq:duality}) is
satisfied. In this case, we can employ this duality transformation,
$\Delta \to 1/\Delta$ and $U \to U/\Delta^2$, to show that Eq.~(\ref{eq:low})
also describes the phase boundary in the limit of \emph{strong} disorder.

As the most generic representative for disorder, we now turn to the case of
`box' disorder ($B$) for which the probability of $\mu_n$ falling at any
point in the range $-W/2 \leq \mu_n \leq W/2$ is equally likely. The
low-energy behavior as shown in the numerical simulation in
Fig.~\ref{fig:disorder}(a) is in good agreement with Eq.~(\ref{eq:low})
(for box disorder, $U = W^2/12$). Eq.~(\ref{eq:specialpoint1}) reveals that
the phase boundary passes through the point $(W,\Delta) = (W_c,1)$, where
$W_c = 4e \approx 10.873$ (box disorder) with $e$ being
the base of the natural logarithm.

A noteworthy find is the observed discontinuity suffered by the phase boundary
as it passes through the random field quantum Ising point $\Delta = 1$
(Fig.~\ref{fig:disorder}(c-d)). To understand its origin, we note that
$\gamma \left( \frac{W}{\sqrt{1-\Delta^2}} \right)$, the effective Lyapunov
exponent that we seek in Eq.~(\ref{eq:phaseboundary}) corresponds to that of
very strong disorder for $\Delta\to 1$. In this
limit, we can use the known form of the normal state Lyapunov exponent for
$W \gg 1$~\cite{disorderreview} in Eq.~(\ref{eq:phaseboundary}), invoke
self-duality and obtain the phase boundary to linear order around $(W,\Delta)
= (4e, 1)$,
\begin{equation} \Delta \approx
\begin{cases}
 \frac{e}{2e^2 + 2}W-\frac{e^2-1}{e^2+1} \ \ \mbox{for } \Delta \le 1, \\
 \frac{e}{2e^2-2}W-\frac{e^2+1}{e^2-1} \ \ \mbox{for } \Delta \ge 1.
\end{cases} \label{eq:ansatz} \end{equation}
As seen in Fig.~\ref{fig:disorder}(c), this result is in reasonable agreement
with numerical simulations. To go further, treating the quantity $\delta =
\frac{1-\Delta}{1+\Delta}$ perturbatively reveals corrections to
Eq.~(\ref{eq:specialpoint1}) yielding $\langle \ln | \mu_n | \rangle = \ln 2
- \left( 1 + \langle 1 / \mu \rangle^2 \right) \frac{\delta}{2} +
\mathcal{O}(\delta^2).$ This shows that the phase boundary is fragile towards
singularities when $\mu_n$ is allowed to come arbitrarily close to zero.
Indeed, our simulations have shown that the discontinuity is absent for
disorder distributions that avoid zero energy. From our extensive simulations
and general insights in the box disorder case, we conclude that a large class
of disorder distributions that cover zero energy give rise to a discontinuity
in the slope of the topological phase boundary at $\Delta = 1$.

Finally, we turn to the specific
case of disorder drawn from a Lorentzian distribution
\begin{equation} f_L(x; W) = \frac{1}{\pi} \frac{W}{x^2 + W^2}.
\end{equation}\
The phase diagram is exactly soluble in this case since the normal state
density of states is known exactly~\cite{lloyd}. The zero-energy Lyapunov
exponent, first obtained by Thouless~\cite{thoulessdos}, takes the
form $\gamma_L(W) = \ln \left( W/2 + \sqrt{1 + W^2/4} \right)$. Once again
invoking Eq.~(\ref{eq:phaseboundary}) and self-duality of the phase diagram
yields a phase boundary
\begin{equation} W =
\begin{cases}
 2 \Delta \ \ \mbox{for } \Delta \le 1, \\
 2 \ \ \mbox{for } \Delta \ge 1.
\end{cases} \label{eq:lorentzian} \end{equation}
This result, as shown in Fig.~\ref{fig:disorder}(c), is in excellent agreement
with numerical simulations. It should be pointed out that the features of this
phase diagram are extremely unusual. For instance, Eq.~(\ref{eq:low}) fails to
hold because the second moment $\langle \mu_n \mu_{n'} \rangle$ is ill-defined
for $f_L$. This example is noteworthy since, for $W > 2$ the system is always
in an N-phase; no amount of $\Delta$ can drive the system into a T-phase.
Studying these examples has shown us, among other features, that typically the
larger the disorder, the more superconductivity is required for Majorana end
modes to exist, and that the topological phase diagram is highly sensitive to
the nature of the disorder distribution.

In conclusion, forging a connection between the normal state localization
properties and the behavior of Majorana end modes has provided us a powerful
means for constructing the complete topological phase diagram for
superconducting wires. Future work would include more extensive utilization
of known Anderson localization results and other physical considerations,
such as interactions and finite temperature.

\emph{Note} -- In the very final stages of preparation of this manuscript, we have been made aware of another work which has some overlap with ours: Ref.~\cite{density}.

We are grateful to I. Gruzberg and V. Shivamoggi for their comments.
For support, we thank the NSF under grant DMR 0644022-CAR (W.D. and S.V.)
and DST, India under grant SR/S2/JCB-44/2010 (D.S.).

\end{document}